\documentclass[12pt,a4paper]{article}

\usepackage{amsmath,amssymb}      
\begin{document}

\title{Static perfect fluids with Pant-Sah equations of state
\thanks{Supported by grants 
FIS2006-05319 (Ministerio de Educaci\'on y Tecnolog\'ia) and 
SA010C0 (Junta de Castillia y Le\'on)}}
\author{Walter Simon\\Facultad de Ciencias, Universidad de Salamanca\\
Plaza de la Merced s/n, E-37008 Salamanca, Spain.\\
              email: walter@usal.es}


\maketitle

\begin{abstract}

We analyze the 3-parameter family of exact, regular, static, spherically
symmetric perfect fluid solutions of Einstein's equations  
(corresponding to a 2-parameter family of equations of state) due to Pant and Sah and "rediscovered" by 
Rosquist and by the present author. Except for the Buchdahl solutions which
are contained as a limiting case, the fluids have finite radius and are physically realistic
for suitable parameter ranges.

The equations of state can be characterized geometrically by the  property that the 
3-metric on the static slices, rescaled conformally with the fourth power 
of any linear function of the norm of the static Killing vector, has constant scalar curvature. 
This local property does not require spherical symmetry; in fact it simplifies the proof of spherical 
symmetry of asymptotically flat solutions which we recall here for the Pant-Sah
equations of state.

We also consider a model in Newtonian theory with analogous geometric and physical properties, 
together with a proof of spherical symmetry of the asymptotically flat solutions.

\end{abstract}

\section{Introduction}
\label{intro}

Solutions of the Tolman-Oppenheimer-Volkoff equation yield quite realistic models for non-rotating stars.
Moreover, the system of the static Einstein equations with perfect fluids also 
provides a testcase for general mathematical techniques and has stimulated their development.
Firstly, group theoretical and Hamiltonian methods for generating solutions have applications to
this system on the local level (see, e.g. \cite{ES,KR1}).
Secondly,  under natural global conditions like asymptotic flatness the system is overdetermined, 
which should lead to spherical symmetry of the solutions for all equations of state (EOS) $\rho(p)$
with $\rho \ge 0$ for $p \ge 0$. This long-standing conjecture has, in essence, been settled recently by 
Masood-ul-Alam \cite{MA1} by using extensions of the techniques of Witten's positive mass theorem
\cite{EW}. Thirdly, as an interesting result on ODEs Rendall and Schmidt \cite{RS} and
Schaudt \cite{US} have proven existence of  1-parameter families of spherically
symmetric, asymptotically flat solutions of the Tolman-Oppenheimer-Volkoff
equation for very general classes of EOS, and for all values of the central pressure and the 
central density for which the EOS is defined.

To illustrate these mathematical results, to study remaining conjectures
(see e.g. \cite{WS1,SYY}) and to make the connection with physics,
it is desirable to have at one's disposal some "exact" model solutions,
preferably with physically realistic equation of state. 
Well suited for these purposes is the two parameter family of Pant-Sah EOS  
(PSEOS) \cite{PS}
\begin{eqnarray} 
\label{psr}
\rho & = & \rho_{-}(1 - \lambda)^{5} + \rho_{+}(1 + \lambda)^{5} \\
\label{psp}
 p & = & \frac{1}{6 \lambda}[\rho_{-}(1 - \lambda)^{6} -
\rho_{+}(1 + \lambda)^{6}] 
\end{eqnarray}
for some constants $\lambda$, $\rho_{-}$ and $\rho_{+}$ with $0<\lambda<1$
and $0\le\rho_+ < \rho_-$. 
This is a parametric representation of solutions of the second order ODE
$I[\rho(p)] \equiv 0$ where  
\begin{equation} 
\label{Ikap}
I[\rho(p)] = \frac{1}{5}\kappa^{2} + 2\kappa + (\rho + p)\frac{d\kappa}{dp} 
\qquad \mbox{with} \quad \kappa = \frac{\rho + p}{\rho +
3p}\frac{d\rho}{dp}.
\end{equation}
Putting  $\rho_+ = 0$ and eliminating $\lambda$ in (\ref{psr}),(\ref{psp}) 
we obtain the one-parameter family of Buchdahl equations of state (BEOS) 
with has the 2-parameter family of Buchdahl solutions \cite{HB1}.
The case $\rho = const.$ is included here as a (degenerate) solution of $I\equiv 0$, 
and it also arises in the limit $\rho_+ \rightarrow \rho_-$, $\lambda
\rightarrow 0$ in (\ref{psr}),(\ref{psp}) (c.f. Sect. 2.2.2).
The general 2-parameter family (\ref{psr}),(\ref{psp}) was considered by Pant and Sah \cite{PS} 
who gave the 3-parameter family of corresponding solutions in terms of elementary functions. 
A decade later, the PSEOS arose in the course of work on a uniqueness proof \cite{BS1,BS2}, 
which lead to the first "rediscovery" of the Pant-Sah solutions \cite{WS2}. 
Moreover, the Pant-Sah solutions also came up in a systematic Hamiltonian approach to 
relativistic perfect fluids by Rosquist \cite{KR1}.

These papers also established the basic properties of the solutions relevant
for their use as stellar models, namely:
\begin{itemize}
\item  The Pant-Sah solutions are regular as long as the central pressure stays bounded
(contrary to the claim by Delgaty and Lake \cite{DL}; see however \cite{KL}
for a correction).
\item All Pant-Sah solutions except for the Buchdahl solutions have a fluid region of finite extent, 
which is obvious from (\ref{psr}), (\ref{psp}) since $\rho > 0$ at $p=0$ iff $\rho_+ > 0$. 
\item The energy density is positive and the pressure is non-negative everywhere, 
and these functions decrease monotonically with the radius.
\item Under suitable restrictions on the parameters, the speed of sound
remains subluminal everywhere \cite{KR2}.
\end{itemize}
Morover, Pant and Sah showed that the parameters can be fitted quite 
well to neutron star data \cite{PS}, while Rosquist \cite{KR2} considered
Pant-Sah solutions as possible "traps" for gravitational waves.

The purpose of this paper is to give a unified description of the Pant-Sah
solutions, to explain their wide range of applicability and to extend it even further.
Apart from the physically relevant properties  mentioned above, we find here the following. 
Depending on the choice of $\rho_+$ and $\rho_-$, the mass-radius relation
(which is a polynomial equation quadratic in the mass) is either monotonic, 
or it exhibits a maximum of the radius only, or a maximum of the mass as well
before it reaches a solution with a singular center.
However, the surface redshift and therefore the quotient mass/radius  
uniquely characterizes a Pant-Sah solution for any given EOS. This implies that the mass-radius curves
can form a single, open "loop" but will not exhibit the "spiral" form typical for degenerate matter 
at extreme densities \cite{HTWW,TM}.
Nevertheless, for suitable $\rho_+$ and $\rho_-$ the mass-radius curve fits remarkably 
well with some quark star models discussed in the last years (see, e.g. \cite{WNR}-\cite{DBDRS}) 
except at extreme densities.

As to the mathematical properties of the Pant-Sah solutions, the key for their
understanding is the {\it "Killing-Yamabe property"}. 
By this term, motivated by the Yamabe problem \cite{LP}, we mean the following: 
{\it For all static solutions of Einstein's equations with a perfect fluid} 
(defined only locally and not necessarily spherically symmetric)
{\it  we require that $g_{ij}^+ = (1 + fV)^4 g_{ij}/16$ is a metric of constant scalar curvature} ${\cal R}_+$
where $g_{ij}$ is the induced metric on the static slices, $V$ is the norm of the Killing vector 
and $f$ is a constant chosen a priori \cite{WS2}.
(If $f\ne 0$, it may be absorbed in $V$ by a suitable scaling of the Killing vector.  
The case $f = 0$ clearly corresponds to fluids with $\rho = const.$).
If the Killing-Yamabe property holds, the field equations {\it imply that} 
$g_{ij}^{-} = (1 - f V)^4 g_{ij}/16$, {\it has constant curvature ${\cal R}_-$
as well, with ${\cal R}_- \ne {\cal R}_+$ in general}.
Together with (\ref{psr}),(\ref{psp}) and  $I\equiv 0$, the Killing-Yamabe
property provides a third alternative 
characterization of the PSEOS,  and the two curvatures are related to the constants in (\ref{psr})
and (\ref{psp}) by ${\cal R}_{\pm} = 512 \pi \rho_{\pm}$.

To understand how the Killing-Yamabe property leads to the "exactness" of the Pant-Sah solutions, we note that 
 spherically symmetric 3-metrics with constant scalar curvature and 
regular centre are "Einstein spaces" (i.e. the Ricci tensor is pure trace).
Such spaces enjoy simple expressions in suitable coordinates, and the same 
applies to the conformal factors defined above.

To sketch the proofs of spherical symmetry of asymptotically flat solutions, 
we need two further generally defined rescalings of the spatial metric, namely 
$g_{ij}^{\ast} =  K(V) g_{ij}$ where the function $K(V)$ is chosen such that 
$g_{ij}^{\ast}$ is flat if $g_{ij}$ is a Pant-Sah solution ($K(V$) is non-unique in
general), and $g_{ij}'= (1 - f^2 V^2)^{4}g_{ij}/16 V^2$.
We then show that for solutions with PSEOS there holds the "$\pm$"- pair of equations 
\begin{equation}
\label{LapR}
\Delta' {\cal R}_{\ast} = \beta_{\pm} {\cal B}_{ij}^{\pm} {\cal B}^{ij}_{\pm}
\end{equation}
where $\Delta'$ is the Laplacian of $g_{ij}'$, ${\cal B}^{+}_{ij}$ and
${\cal B}^{-}_{ij}$ are the trace-free parts of the Ricci-tensors of $g_{ij}^{+}$
and $g_{ij}^{-}$ respectively, and $\beta_{+}$ and $\beta_-$ are non-positive functions.

To show uniqueness of the Buchdahl solutions for the BEOS (${\cal R}_{+} = 0$, ${\cal R}_- \ne 0$), 
one first shows that all asymptotically flat solutions must extend to infinity
\cite {WS1,WS3,MH}. Hence for a Killing vector normalized such that
$V\rightarrow 1$ at infinity, we can choose $f = 1$ and  $K(V) = (1 + V)^4/16$ in the definitions
above. Then there are two alternative ways to continue \cite{BS1}.  The first one consists of 
noting that $g_{ij}^{\ast} = g_{ij}^+$  is asymptotically flat with vanishing mass. 
Hence the positive mass theorem \cite{EW,SY} implies that these metrics are flat and $(V,g_{ij})$ is
a Buchdahl solution.
Alternatively, we can integrate the "minus" version of (\ref{LapR}) over the static slice. 
By the divergence theorem and by asymptotic flatness,
${\cal B}_{ij}^- = 0$, i.e. $g_{ij}^-$ is an Einstein space, which again yields
a Buchdahl solution.

In the generic case ${\cal R}_{+} \ne 0$, the divergence theorem alone 
applied to (\ref{LapR}) is insuffient for a proof as ${\cal R}_{\ast}$ 
cannot be made $C^1$ on the fluid boundary in general. 
However, by employing a suitable elliptic identity in the vacuum
region as well, the maximum principle now yields ${\cal R}_{\ast} \ge 0$.
Then the positive mass theorem leads to the required conclusion.

The positive mass theorem combined with Equ. (\ref{LapR}) has been been employed earlier
in proofs 
of spherical symmetry in the cases of fluids with constant density \cite{Lind} and "near constant density"  \cite{MA2}.  
Moreover, there are generalizations of (\ref{LapR}) for non-Killing-Yamabe EOS, for fluids which satisfy $I\le 0$, 
which again give uniqueness \cite{BS2,LM}. The general proof of spherical symmetry by Masood-ul-Alam
\cite{MA1} involves modified Witten spinor identities and integral versions of generalizations of (\ref{LapR}).

Jointly with the PSEOS we will consider here a model in Newtonian theory
characterized by the following counterpart of the Killing-Yamabe property: 
 We require that {\it a conformal rescaling of flat space with 
$(\bar v - v)^4/16$,} where $v$ is the Newtonian potential (not necessarily spherically
symmetric) and $\bar v$ a constant, {\it is a metric of constant scalar
curvature} ${\cal R}_-$. This leads to the 2-parameter family of equations of state 
\begin{equation}
\label{Neos}
p = \frac{1}{6} \left( \rho_{-}^{- \frac{1}{5}}~ \rho^{\frac{6}{5}} - \rho_+ \right)
\end{equation}
where  ${\cal R}_- = 512 \pi \rho_-$ and $\rho_+$ is another constant which has here no obvious relation to curvature.  
We will refer to (\ref{Neos}) as "the Newtonian equation(s) of state" (NEOS).
For $\rho_+ = 0$ the NEOS are polytropes of index 5 which are analogous to the BEOS. 
The general NEOS and the corresponding solutions may be considered as "Newtonian limits" of the PSEOS and the
Pant-Sah solutions, with similar properties for low density and pressure.
As to uniqueness proofs for asymptotically flat solutions with the NEOS, 
there is available some sort of counterpart of (\ref{LapR}), and the positive mass theorem
has to be substituted here by the "virial theorem".

This paper is organized as follows. In Sect. 2 we give the field
equations in the Newtonian and in the relativistic case and introduce our models. 
In Sect.3 we rederive the spherically symmetric solutions and discuss 
their main properties, in particular the mass-radius curves.
The Section 4 we prove spherical symmetry of asymptotically flat solutions
with the NEOS and the PSEOS. The 
Appendix contains general material on conformal rescalings of metrics
and on spaces of constant curvature.

\section{The Field Equations}

Our description of the Newtonian and the relativistic fluids will be as close as
possible. For simplicity we will use identical symbols 
($g_{ij}$, $g_{ij}^-$, $\nabla_i$, ${\cal R}$, ${\cal R}^-$...)
for analogous quantities but with different formal definitions, depending on
the context.

We denote by $\cal F$ the fluid region, which we assume to be open and connected, 
and which may extend to infinity. $\cal V$ is the open vacuum region
(which may be empty) and $\partial {\cal F} = \partial {\cal V}$ is the 
{\it common} boundary (i.e. infinity is not included in ${\partial \cal V}$ ). 
This redundant terminology is useful to describe the matching.
When $\cal F$ is spherically symmetric it is called "star".

\subsection{ Newtonian Fluids}

\subsubsection{General properties}

We consider as Newtonian model  $\cal F \cup \partial {\cal F} \cup \cal V$  a manifold 
$({\cal M}, g_{ij})$ with a flat metric $g_{ij}$.  The potential function $v$ is assumed
to be smooth in ${\cal F}$ and ${\cal V}$, $C^{1,1}$ at $\partial {\cal F}$,
negative everywhere and $v\rightarrow 0$ at infinity. 
For smooth density $\rho(x^i)$ and pressure functions $p(x^i)$ in ${\cal F}$, 
with $p \rightarrow 0$ at $\partial {\cal F}$, Newton's and Euler's equations read  
\begin{eqnarray}
\label{Poi}
\Delta v & = & 4\pi \rho \\
\label{Eul}
\nabla_i p & = & - \rho \nabla_i v
\end{eqnarray}
where $\nabla_i$ and $\Delta = \nabla_i \nabla^i$  denote the gradient and the Laplacian of flat space. 
(Indices are moved with $g_{ij}$ and its inverse $g^{ij}$).

A general EOS is of the form $H(\rho,p) = 0$. (If possible we choose
$H(\rho,p) = p - p(\rho)$). $H(\rho,p)$ should be defined in the intervals $\rho \in
[\rho_s,\infty)$ and $p \in [0,\infty)$ with $\rho_s \ge 0$
and smooth in the intervals $\rho \in (\rho_s,\infty)$ and $p \in (0,\infty)$.
Using the EOS, we can write $p$ and $\rho$ as smooth functions $p(v)$ and
$\rho(v)$ and Euler's equation (\ref{Eul}) as $dp/dv = - \rho$.

To recall the matching conditions, we note that, from the above requirements, the metric induced on 
$\partial {\cal F}$ is $C^{1,1}$ and the mean curvature of $\partial {\cal F}$ is continuous. 
We now write these conditions in terms of the quantity $w=\nabla_i v \nabla^i v$.
(The gradient always acts only on the subsequent argument, i.e. $w=(\nabla_i v)( \nabla^i v)$).
From (\ref{Poi}) it follows that the quantity in brackets on the l.h. side of 
\begin{equation}
\label{Ngm}
 \left[w^{-1} \nabla^i v \nabla_i w - 8 \pi \rho \right]_{\Rightarrow \partial{\cal F}}  
=   \left[w^{-1}\nabla^i v \nabla_i w \right]_{\partial{\cal V} \Leftarrow} 
\end{equation}  
is continuous at the surface.  Hence (\ref{Ngm}) must hold, where
"$\Rightarrow \partial{\cal F}$" and "$\partial{\cal V} \Leftarrow$" denote
the approach to the boundary from the fluid and the vacuum sides, respectively.   

Generalizations to several disconnected "matching surfaces" are
trival and will not be considered here.

To formulate the asymptotic properties we consider an "end" 
${\cal M}^{\infty} = {\cal M} \setminus \{ \mbox{a compact set} \}$. 
We assume that, for some $\epsilon > 0$
\begin{equation}
\label{NM}
v = - \frac{M}{r} + O(\frac{1}{r^{1+\epsilon}}), \quad 
\partial_i v = - \partial_i \frac{M}{r} + O(\frac{1}{r^{2+\epsilon}}) \quad
\partial_i \partial_j v = - \partial_i \partial_j \frac{M}{r} +
O(\frac{1}{r^{3+\epsilon}})
\end{equation}
where $M$ is the mass.
With (\ref{Poi}) and (\ref{Eul}) this implies that
\begin{equation}
\label{Nas}
\rho = O(\frac{1}{r^{3+\epsilon}}), \qquad p = O(\frac{1}{r^{4+\epsilon}}).
\end{equation}
A more natural but involved precedure is to derive the falloff conditions
of the potential $v$ and of $p$ only from the falloff of $\rho$ 
(c.f. \cite{WS1}).

 \subsubsection{The Newtonian equation of state}

To introduce our model it is useful to rescale the Euclidean metric by the
fourth power of a linear function of the Newtonian potential, i.e.
we define $g_{ij}^- = g_{ij} (\bar v - v)^4/16$ for some constant $\bar v \ge 0$. 

We use the general formula $(\ref{csc})$ for conformal rescalings with 
$\wp_{ij}$ flat, $\Phi = (\bar v - v)/2$, and hence $\widetilde \wp_{ij} =
g_{ij}^-$. Together with the field equation (\ref{Poi}), we find that
\begin{equation}
\label{conf}
 {\cal R}_- (\bar v -  v)^5 = 512\pi \rho
\end{equation}
where ${\cal R}_-$ is the scalar curvature of $g_{ij}^-$.
We now determine the NEOS by requiring that ${\cal R}_- = const.$. 
Introducing $\rho_-$ by ${\cal R}_- = 512 \pi \rho_- $ and another constant
$\rho_+ $, eqs. (\ref{Eul}) and (\ref{conf}) yield
\begin{equation}
\label{rhop}
 \rho = \rho_- (\bar v - v)^5 \qquad p  =  \frac{1}{6}[\rho_- (\bar v - v)^6 - \rho_+].
\end{equation}
We require that $\rho_+ \ge 0$ and $\rho_- > 0$.
Eliminating the potential we obtain the NEOS equ. (\ref{Neos}).
In terms of the variables $p/\rho_-$ and $\rho/\rho_-$ and  $\tau = (\rho_+/\rho_-)^{1/6}$
this equation reads
$p/\rho_- = \frac{1}{6} \left[ (\rho/\rho_-)^{\frac{6}{5}} - \tau^6 \right]$. 
This means that we have singled out $\rho_-$ as a "scaling" parameter while 
$\tau \in [0,\infty]$ plays a more "essential" role. 
(This terminology mainly serves to simplify the analysis of static spherically symmetric
solutions in Sect. 3. Both parameters have direct physical significance, as follows from 
(\ref{Nsurf}) below. On the other hand, from the dynamical system point of
view, both parameters can be considered as "scaling" except in the case $\rho =const.$
(c.f. \cite{HU,HRU}). 

\begin{figure}[h!]
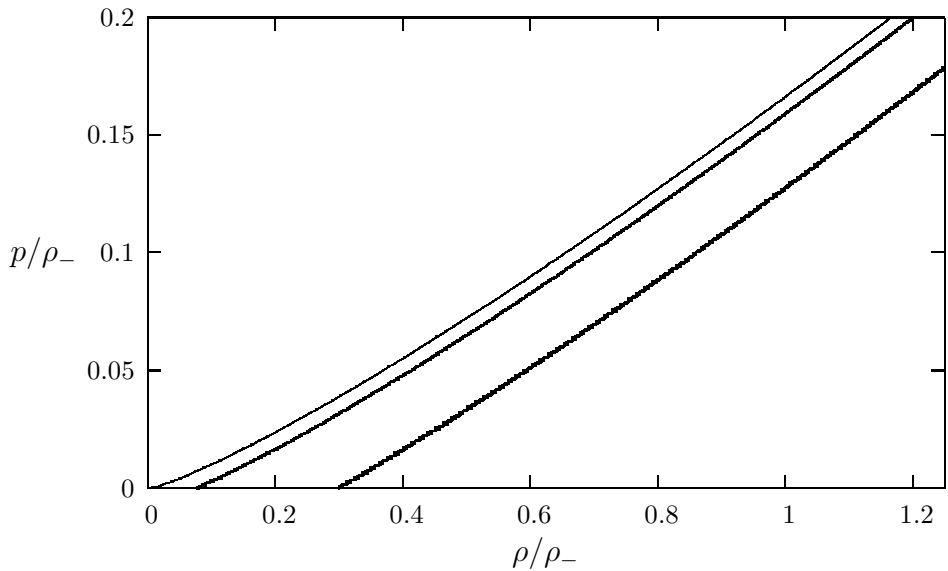

\setlength{\unitlength}{0.240900pt}
\ifx\plotpoint\undefined\newsavebox{\plotpoint}\fi
\sbox{\plotpoint}{\rule[-0.200pt]{0.400pt}{0.400pt}}%

\caption{The equation of state (\ref{Neos}) for the values $\tau = 0.382$ (thin line),
$\tau = 0.6$ (medium) and $\tau= 0.786$ (thick line).}
\label{NEOS}    
\end{figure}

Fig. (\ref{NEOS}) shows the NEOS for the values
$\tau = (3 - \sqrt{5})/2 \approx 0.382$, $\tau = 0.6$ and 
$\tau = [(\sqrt{5} - 1)/2]^{1/2} \approx 0.786$. 
(These particular values play a role in Relativity and are chosen here for
comparison).

The speed of sound $C$ defined by the first equation of (\ref{ss}) takes the
simple form 
\begin{equation}
\label{ss}
C^2  = \frac{dp}{d\rho} = \frac{1}{5}\left( \frac{\rho}{\rho_-} \right)^{\frac{1}{5}}
= \frac{1}{5}(\bar v - v)
\end{equation}
in terms of the potential.
We recall that $\bar v$ was taken to be positive, hence $dp/d\rho > 0$ and (\ref{ss}) makes sense. 

At the surface where the pressure is zero the potential, the density and the speed of 
sound take the values 

\begin{equation}
\label{Nsurf}
v_s = \bar v - \tau, \qquad \rho_s=  \tau^{-1} \rho_{+}, \qquad C_s^2 = \frac{1}{5}\tau.
\end{equation}

Note that $\rho_s$ and $C_s$ are determined by the equation of state alone, 
as opposed to $v_s$ and $\bar v$ which will be used in Sect. 3.1 to parametrize the spherically symmetric solutions. 
 The polytrope of index 5, for which the solutions extend to infinity, arises from the equations above 
as the special case $\rho_+ = \rho_s = 0 = v_s = \bar v$. The corresponding
curve would pass through the origin in Fig. (\ref{NEOS}), very close to the curve for $\tau = 0.382$. 

\subsection{Relativistic Fluids}

\subsubsection{General properties}

We consider static spacetimes of the from 
$\mathbb{R} \times {\cal M} = \mathbb{R} \times 
\left({\cal F} \cup \partial {\cal F} \cup {\cal V} \right)$ 
with metric 
\begin{equation} 
\label{met}
ds^{2} = - V^{2}dt^{2} + g_{ij}dx^{i}dx^{j}
\end{equation}
where $V(x^i)$ and $g_{ij}(x^i)$ are smooth on ${\cal F}$ and ${\cal V}$ and $C^{1,1}$ at $\partial {\cal F}$. 
Moreover, $0 < V < 1$ on ${\cal M}$ and $V \rightarrow 1$ at infinity. 
On ${\cal F}$ we consider smooth density and pressure functions 
$\rho(x^i)$, $p(x^i)$, with $p\rightarrow 0$ on $\partial {\cal F}$, 
in terms of which Einstein's and Euler's equations read  
\begin{eqnarray} 
\label{Alb}
\Delta V & = & 4 \pi V (\rho + 3p) \\
\label{Ein}
{\cal R}_{ij} & = & V^{-1}\nabla_{i}\nabla_{j}V + 4 \pi(\rho - p)g_{ij}\\
\label{Bia}
\nabla_i p & = & - V^{-1} (\rho + p) \nabla_i V.
\end{eqnarray}
The gradient $\nabla_i$, the Laplacian $\Delta = \nabla_{i}\nabla^{i}$ and the Ricci
tensor ${\cal R}_{ij}$ now refer to $g_{ij}$. 
As well known the Euler equation (\ref{Bia}) is a consequence of the Bianchi identity for
${\cal R}_{ij}$.

A general equation of state $H(\rho,p) = 0$, in particular with $H(\rho,p) = p -
\rho(p)$, should be defined in $\rho \in [\rho_s,\infty)$ and $p \in [0,\infty)$ 
with $\rho_s \ge 0$ and smooth in the intervals $\rho \in (\rho_s,\infty)$ and $p \in (0,\infty)$. 
Euler's equation (\ref{Bia}) together with the equation of state imply that there are smooth 
functions $p(V)$ and $\rho(V)$, and Euler's equation becomes $dp/dV = - V^{-1}(\rho + p)$.

In analogy with the Newtonian case, the metric induced on $\partial {\cal F}$ is $C^{1,1}$ 
and the mean curvature of $\partial {\cal F}$ is continuous. 
In terms of the quantity $W=\nabla_i V \nabla^i V$, the matching conditions
together with equ. (\ref{Alb}) imply that the quantity in brackets on the l.h. side of 
\begin{equation}
\label{Egm}
 \left[W^{-1} \nabla^i V \nabla_i W - 8 \pi V \rho \right]_{\Rightarrow \partial{\cal F}}  
=   \left[W^{-1}\nabla^i V \nabla_i W \right]_{\partial{\cal V} \Leftarrow} 
\end{equation}  
is continuous at $\partial {\cal F}$ and hence (\ref{Egm}) holds. 

To formulate the asymptotic properties we consider an "end" 
${\cal M}^{\infty} = {\cal M} \setminus \{ \mbox{a compact set} \}$. 
We require that, for some $\epsilon > 0$
\begin{eqnarray}
\label{EM}
 V   =  1 - \frac{M}{r} + O(\frac{1}{r^{1+\epsilon}}), \quad
 \partial_i V  & = & - \partial_i \frac{M}{r} + O(\frac{1}{r^{2+\epsilon}}), \nonumber \\  
\partial_i \partial_j V  & = & - \partial_i \partial_j  \frac{M}{r} +
O(\frac{1}{r^{3+\epsilon}}), \\ 
\label{Eg}
g_{ij}  = (1 + \frac{2M}{r})\delta_{ij} + O(\frac{1}{r^{1+\epsilon}}), \quad
\partial_k g_{ij} & = &\partial_k \frac{2M}{r}\delta_{ij} +
O(\frac{1}{r^{2+\epsilon}})
\nonumber \\
\partial_k \partial_l g_{ij} & = & \partial_k \partial_l  \frac{2M}{r} \delta_{ij} +
O(\frac{1}{r^{3+\epsilon}})
\end{eqnarray}
in suitable coordinates, where $M$ is the mass.
Equs.
 (\ref{Alb}) and (\ref{Bia}) together with the decay conditions (\ref{EM}) and (\ref{Eg}) imply that
\begin{equation}
\label{Eas}
\rho = O(\frac{1}{r^{3+\epsilon}}), \qquad p = O(\frac{1}{r^{4+\epsilon}}).
\end{equation}
Here the falloff conditions of the potential $V$ could also be derived from
some weak falloff conditions of $g_{ij}$, and $\rho$ and  $p$
\cite{WS1}. 
Clearly a substantial refinement of all asymptotic conditions is possible if 
 ${\cal M}^{\infty}$ is vacuum, c.f. \cite{KM}. 

\subsubsection{The Pant-Sah equation of state}

We now introduce conformal rescalings of the spatial metric of the form 
$g_{ij}^{\pm} = g_{ij} (1 \pm fV)^4/16$ for some constant $f$ which we take to be
non-negative (this just fixes the notation), 
and we restrict ourselves to the range $V < 1/f$.
While any  $f > 0$  could be absorbed into $V$ by rescaling the static Killing
field, we have already fixed the scaling above by requiring $V \rightarrow 1$ at infinity, 
which is why the extra constant $f$ will persist here in general.  
We now use the standard formula (\ref{csc}) with  $\wp = g_{ij}$, $\Phi = (1 \pm fV)/2$ so
that $\widetilde \wp_{ij} = g^{\pm}_{ij}$. Together with the field equations (\ref{Alb}) and 
(\ref{Ein}) this gives
\begin{equation}
\label{Rpm}
\frac{1}{128} {\cal R}_{\pm}(1 \pm fV)^5 = - (\Delta - \frac{1}{8}{\cal R})(1 \pm fV) = 2\pi[\rho(1 \mp fV) \mp 6fpV]
\end{equation}  
where ${\cal R}$ and ${\cal R}_{\pm}$ are the scalar curvatures of $g_{ij}$ and
$g_{ij}^{\pm}$, respectively. By differentiating (\ref{Rpm}) with respect to $V$ we obtain
\begin{equation} 
\label{dRdV}
\frac{d {\cal R}_{\pm}}{dV} = \frac{2560 \pi (\rho + 3p)}{(1 \pm fV)^{6}}
\left[f^2 V - \frac{\kappa}{10 V} \left(1 - f^2 V^2\right) \right] 
\end{equation}
where $\kappa$ has been introduced in (\ref{Ikap}).

We now implement the "Killing Yamabe property" defined in the introduction
by requiring that at least one of the curvatures ${\cal R}_{+}$ and ${\cal R}_{-}$ is constant. 
This implies that the quantity in brackets in (\ref{dRdV})  vanishes, i.e.,
\begin{equation}
 \label{kap} 
\kappa = \frac{10 f^{2}V^{2}}{(1 - f^{2}V^{2})}
\end{equation} 
and therefore the other scalar curvature is necessarily constant as well.
Using this in (\ref{Rpm}) and setting ${\cal R}_{\pm} = 512 \pi \rho_{\pm}$ 
we obtain 
\begin{eqnarray} 
\label{rV}
\rho & = & \rho_{-}(1 - fV)^{5} + \rho_{+}(1 + fV)^{5}, \\
\label{pV}
 p & = & \frac{1}{6 fV}[\rho_{-}(1 - fV)^{6} - \rho_{+}(1 + fV)^{6}] 
\end{eqnarray}
which yields the parametric form (\ref{psr}), (\ref{psp}) of the PSEOS 
when we set $fV = \lambda$. Positivity of the pressure now requires that we
restrict ourselves to $0 \le \rho_+ < \rho_- < \infty$. 
Note that Equ. (\ref{kap}) implies that $d\rho/dp$ takes on all real positive values for the 
allowed range $0< fV < 1$ of $V$.

The case $\rho = const.$ is included in (\ref{rV}) and (\ref{pV}) in the
limit $\rho_+ \rightarrow \rho_-$ and $fV \rightarrow 0$. 
To see this we expand (\ref{rV}) and (\ref{pV}) in $fV$, 
\begin{equation} 
\label{rhoc}
\rho = \rho_+ + \rho_- + O(fV), \quad 
p =  \frac{\rho_- - \rho_+}{6 f V} - (\rho_+ + \rho_-) + O(fV).
\end{equation}
Without the terms of order $fV$, this is a solution of the Euler equation (\ref{Bia}) 
for constant density $\rho_+ + \rho_-$. 
Now the limit $\rho_+ \rightarrow \rho_-$ and $fV \rightarrow 0$ 
has to be taken in such a way that $p$ stays regular and non-negative
(we skip mathematical subtleties).

\begin{figure}[h!]
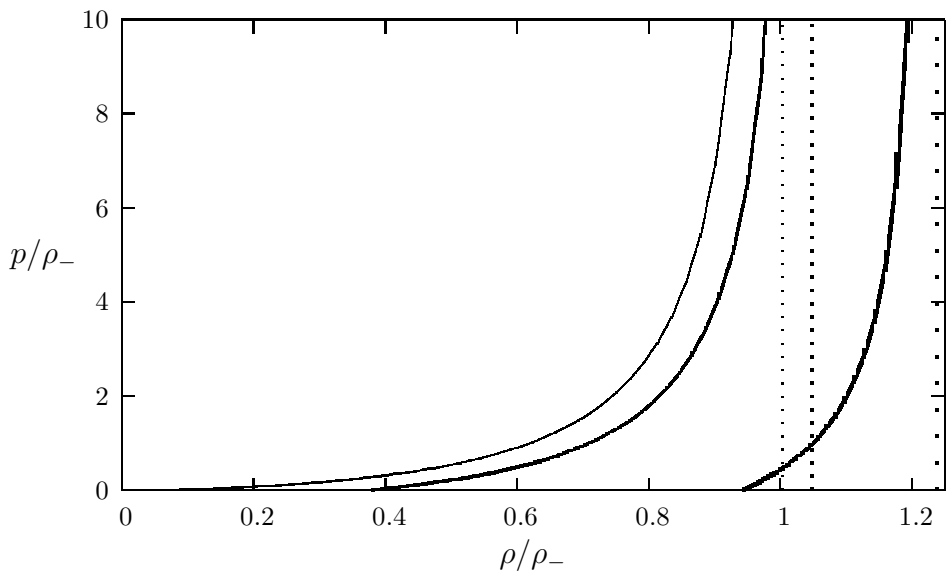


\setlength{\unitlength}{0.240900pt}
\ifx\plotpoint\undefined\newsavebox{\plotpoint}\fi
\sbox{\plotpoint}{\rule[-0.200pt]{0.400pt}{0.400pt}}%

\caption{The Pant-Sah equation of state (\ref{psr}), (\ref{psp}) for the values $\tau= 0.382$ (thin solid line)
$\tau = 0.6$ (medium solid) and $\tau = 0.786$ (thick solid line).
The dotted lines indicate the respective limits of $\rho/\rho_-$ for $p/\rho_- \rightarrow
\infty$.}
\label{PSEOS}      
\end{figure}

We define $\tau = (\rho_+/\rho_-)^{\frac{1}{6}}$ which, in contrast to the Newtonian case, 
is now restricted to be less than $1$.
We draw the PSEOS in Fig. (\ref{PSEOS}) in terms of the rescaled variables $p/\rho_-$, 
$\rho/\rho_-$ and for the same values of $\tau$ as chosen for the NEOS in Fig. (\ref{NEOS}), 
namely $\tau = (3 - \sqrt{5})/2 \approx 0.382$, $\tau = 0.6$ and 
$\tau = [(\sqrt{5} - 1)/2]^{1/2} \approx 0.786$. 

At first sight the PSEOS looks very different from the Newtonian model, 
Fig. (\ref{NEOS}). In fact, in contrast to the latter, 
the density now stays bounded and tends to $\rho_+ + \rho_-$ as the pressure goes to infinity
(which happens for $fV \rightarrow 0$). 
This means that for high pressures the PSEOS first violates the energy
conditions, and finally always becomes infinitely "stiff".
Note however that Fig. (\ref{NEOS}) and Fig. (\ref{PSEOS}) have a very
different scale in the $p/\rho_-$ direction. For small $p/\rho_-$ and small
$\rho/\rho_-$ we can still consider the EOS (\ref{Neos}) as Newtonian limit of the
PSEOS (\ref{rV}) and (\ref{pV}).

At the surface, the quantity $\kappa$ as defined by (\ref{Ikap}) is related
to the speed of sound by  $C_s^2 = dp/d\rho|_s = \kappa_s^{-1}$.
In analogy with (\ref{Nsurf}) we now determine the surface potential,
the surface density and $C_s$ from (\ref{kap}),(\ref{rV}) and (\ref{pV}) as

\begin{equation}
\label{Esurf}
fV_s  = \frac{1 - \tau}{1 + \tau}, \qquad \rho_s = \frac{32\rho_- \tau^5}{(1 +
\tau)^4}, \qquad C_s^2 = \frac{2\tau}{5 (1 - \tau)^2}.
\end{equation}

Since $V_s < 1$, $f$ is bounded from below by $f > (1 - \tau)/(1 + \tau)$.

Again $\rho_s$ and $C_s$ are determined by the EOS alone whereas one of $f$ or $V_s$ 
can be used to parametrize the solutions. If $\tau > (6 - \sqrt{11})/5
\approx 0.537$, then $C_s > 1$, 
i.e. the speed of sound exceeds the speed of light already at the surface.
This applies in particular to the curves for  $\tau = 0.6$ and $\tau =
0.786$ of Fig. (\ref{PSEOS}).
But $\tau < (6 - \sqrt{11})/5$ implies $C_s < 1$, and for
sufficiently "small" spherical stars $C < 1$ then holds up to the center.
An example is the thin line  $\tau = 0.382$ of Fig. (\ref{PSEOS}).
On the other hand, for all $\tau$, $C > 1$ at the centre if the star is
sufficiently large due to the stiffness of the PSEOS at high pressures.
The size limits for the star follow from Fig.1 in \cite{KR2}, and
they could also be determined from the results of Sect. 3.2.

While for our purposes the parametric form (\ref{rV}) and (\ref{pV})
suffices as EOS, the latter can also be 
displayed in closed form. 
We first consider the BEOS  ($\rho_+ = 0$) which reads
\begin{equation}
p = \frac{\rho^{6/5}}{6 (\rho_{-}^{1/5} - \rho^{1/5})},
\end{equation}
and holds for $\rho<\rho_{-}$.
In the general case $\rho_+ > 0$ it is clearly simplest to eliminate one of $\rho_-$ or
$\rho_+$, and to interpret the other one, together with $\lambda = fV$ as parameters
of the PSEOS.
However, in view of the geometric interpretation of  $\rho_-$ and $\rho_+$, 
and in view of the "symmetric form" of equations
(\ref{rV}) and (\ref{pV}), it is more natural to eliminate $\lambda = fV$. 
To do so we note that the following linear combination of equations
(\ref{rV}) and (\ref{pV})
\begin{eqnarray}
\label{poly4}
\lefteqn{\frac{1}{20 \rho_-} [\rho(1 + \lambda) + 6 p\lambda]
+ \frac{1}{20 \rho_+} [\rho(1 - \lambda) - 6 p\lambda]  = {} } \nonumber\\
 &  & {} = \frac{1}{10}(1 - \lambda)^5 +  \frac{1}{10}(1 + \lambda)^5 =
\lambda^4 + 2\lambda^2 + \frac{1}{5}  
\end{eqnarray}
gives a polynomial equation of fourth order in $\lambda$ which can be solved algebraically by a 
standard procedure.
Alternatively, we can use (\ref{poly4}) to eliminate the fifth and fourth order terms in
(\ref{rV}) which leaves us with the polynomial equation
\begin{equation}
\label{poly3}
\lambda^3 + \frac{\nu_-}{5} (\rho + 6p) \lambda^2 + 
\frac{3}{5}[1 + 2\nu_+ (\rho + 5p)] \lambda - \nu_- \rho = 0
\end{equation}
of third order, with $32 \nu_{\pm} = \rho_+^{-1} \pm \rho_-^{-1}$. Solving either (\ref{poly4}) 
or (\ref{poly3}) for $\lambda = \lambda(\rho,p,\rho_+,\rho_-)$ and putting this back again into 
(\ref{rV}) or (\ref{pV}) gives the PSEOS in closed form $H(\rho,p,\rho_+,\rho_-) = 0$.
The function $H$ is elementary but involved  and will not be displayed here.

\section{The Spherically Symmetric Solutions}

We now determine the spherically symmetric solutions corresponding to the two-parameter 
families of NEOS (\ref{Neos}) and the PSEOS (\ref{psr}) and (\ref{psp}),
making use of the formulas in the Appendix.
By the general theorem \cite{RS} there exist 1-parameter families of such
solutions in either case, for all values of the central pressure.
We will in particular determine the physically relevant parameters mass $M$ and radius
$R$. As well known and easy to see from the definitions and the field equations, 
families of static, spherically symmetric fluid solutions are always invariant under the scaling  
\begin{equation}
\rho \rightarrow \gamma^2 \rho, \qquad p \rightarrow \gamma^2 p,
\qquad M \rightarrow \gamma^{-1} M,  \qquad  R \rightarrow \gamma^{-1} R.
\end{equation}
for any $\gamma > 0$. For our families of solutions this means that one of the three 
parameters is "trivial" in this sense. 
In Sects. 3.1.2 and 3.2.3 we will therefore use scale invariant variables $\widehat M$ and 
$\widehat R$ defined by
\begin{equation}
\label{resc}
\widehat M = \sqrt{\frac{8\pi \rho_s}{3}}M,  \qquad  \widehat R = \sqrt{\frac{8\pi \rho_s}{3}} R,
\end{equation}
where $\rho_s$ is the surface density. Note that the latter is given in terms of $\rho_-$ and 
$\rho_+$ in Newtonian theory by (\ref{Nsurf}) but in Relativity by (\ref{Esurf}).  

\subsection{The Newtonian Solutions}

\subsubsection{The matching}

Using Lemma A.3. of the Appendix with $\wp^+_{ij}$ flat, $\Phi = (\bar v - v)/2$, 
and ${\cal R}_- = 512\pi\rho_-$ we can write the spherically symmetric solutions
(\ref{Phi}) of (\ref{conf}) as
\begin{equation}
\label{vmu}
\bar v - v =  2 \mu  \sqrt{\frac{1}{1 + \frac{64 \pi}{3}\mu^4 \rho_- r^2}}.
\end{equation}
It remains to eliminate one of the  constants $\bar v$ and $\mu$ by global conditions.
In the case $\rho_+ = 0$ the NEOS becomes the polytrope of index 5. 
It follows from (\ref{PSvir}) that $ \bar v = 0$ which means that ${\cal F}$ extends to 
infinity and (\ref{vmu}) is valid for all $r$. 
The solutions can be conveniently parametrized by their mass $M$ defined in (\ref{NM}). 

In the case $\rho_+ > 0$, it is simplest to parametrize the solutions in terms 
of $\bar v$ which is related to the surface potential by (\ref{Nsurf}).  
To get $\mu$ it suffices to use that $v \in C^1$ which implies that 
\begin{equation}
\label{Nmcoo}
 \left. \frac{dv}{dr} \right|_{\Rightarrow {\cal F}} =  
 \left. \frac{dv}{dr} \right|_{{\cal V} \Leftarrow} =  - \frac{v_s}{R}.
\end{equation}
where $R = r|_s$ is the radius of the star.

Using (\ref{vmu}) and (\ref{Nsurf}) and recalling that $\bar v$ was assumed
non-negative, it follows that $\mu$ can be expressed as
\begin{equation}
\label{Nmu}
 \mu = \left\{ \begin{array}{r@{\quad \mbox{for} \quad}l} 
 {\displaystyle \frac{1}{M} \sqrt{ \frac{3}{16 \pi \rho_-} } } & \rho_+ = 0, \\
 {\displaystyle \frac{\tau}{2} \sqrt{\frac{\tau}{\bar v} } } & \rho_+ > 0  
\end{array} \right.
\end{equation}
and we can write (\ref{vmu}) as
\begin{equation}
\label{vex}
 v = \left\{ \begin{array}{r@{\quad \mbox{for}\quad}l} 
 {\displaystyle - \frac{M}{\sqrt{\frac{4\pi}{3} \rho_- M^4 + r^2}}  }  & \rho_+ = 0, \\
 {\displaystyle \bar v - \tau \sqrt{\frac{\tau \bar v}{\bar v^2 + \frac{4\pi}{3} \rho_+ r^2}} }
 & \rho_+ > 0.
  \end{array} \right.
\end{equation}

Note that $M$ can take any value $M \in [0,\infty)$, and
the allowed values for the other parameters are 
$\bar v \in [0,\tau]$ or $v_s \in [-\tau,0]$. 
 
For all $\rho_+$ and $\rho_-$ the density, 
the pressure and the speed of sound follow from (\ref{rhop}) and (\ref{ss}); 
they are monotonic functions of $r$. 
For $\rho_+ > 0$ the central density $\rho_c$, the central pressure $p_c$ 
and the speed of sound at the center $C_c$ take the values
\begin{equation}
\rho_c =  \frac{\rho_+ \tau}{\bar v^2}\sqrt{\frac{\tau}{\bar v}},  \qquad 
p_c = \frac{\rho_+}{6} \left( \frac{\tau^3}{\bar v^3} - 1 \right), \qquad 
C_c^2 = \frac{\tau}{5} \sqrt{\frac{\tau}{\bar v}}.
\end{equation}
These quantities diverge as the parameter $\bar v$ goes to zero. 

Instead of the coordinate expressions (\ref{vmu}) - (\ref{vex}) the matching and the solutions
can be described in a "covariant" manner in terms of $w = \nabla_i v \nabla^i v$  
which is a function of $v$ in the spherically symmetry case.
In particular we have $w= M^{-2} v^4$ in the vacuum region.
To determine $w$ for the spherically symmetric solutions characterized by ${\cal R}_- = 0$, 
we use Lemma A.2 of the Appendix which shows that $g_{ij}^- = g_{ij} (\bar v - v)^4 /16$ 
are spaces of constant curvature. With the general formula (\ref{cri}) this yields
\begin{equation}
0 = (\bar v - v)^2 {\cal B}_{ij}^- = 2 {\cal C}[(\bar v - v) \nabla_i\nabla_j v + 3 \nabla_i v \nabla_j v] 
\end{equation}
Contracting this equation with $\nabla^i v \nabla^j v$ and using (\ref{Poi})
and (\ref{rhop}) gives
\begin{equation}
\frac{d}{dv} \left[ \frac{w}{(\bar v - v)^4} \right] = \frac{8\pi}{3}\rho_- (\bar v - v).
\end{equation}
This has the solution 
\begin{equation}
\label{wv}
w = \frac{4 \pi}{3} \rho_- (\bar v - v)^4 
\left[\sigma^2 -  (\bar v - v)^2 \right]
\end{equation}
for some constant $\sigma$ which has to be determined by global conditions.

From the exterior form $w = M^{-2} v^4$ and from the matching conditions (\ref{Ngm})
we obtain 
\begin{equation}
\label{Nsm}
 \left[ \frac{d w}{d v} - 8 \pi \rho
\right]_{\Rightarrow \partial {\cal F}} = \left[ \frac{d w}{d v} \right]_{\partial {\cal V} \Leftarrow} 
=  \frac{4 w_s}{v_s}.
\end{equation}  

Using the asymptotic property (\ref{NM}) for $\rho_+ = 0$ and (\ref{Nsm})
for $\rho_+ > 0$ we find that 

\begin{equation}
\label{sigma}
 \sigma^{2} = \left\{ \begin{array}{r@{\quad \mbox{for} \quad}l} 
 {\displaystyle \frac{3}{16 \pi \rho_- M^2} } & \rho_+ = 0, \\
 {\displaystyle \frac{\tau^3}{\bar v} } & \rho_+ > 0.  \end{array} \right.
\end{equation}

Alternatively, equ. (\ref{wv}) can of course be checked directly from (\ref{vex}).
In particular the value of $\sigma$ follows from (\ref{Nmu}) or vice versa.

\subsubsection{The Mass-Radius relation}

To determine mass and radius we take equ. (\ref{vex}) or the first of
(\ref{Nsm}) and (\ref{wv}) at the surface and use $v_s = -M/R$. 

In terms of the rescaled variables (\ref{resc}) this gives
\begin{equation}
\widehat R^2   =  \frac{2 \bar v}{\tau} (\tau - \bar v), \qquad
\widehat M^2   =  \frac{2 \bar v}{\tau} (\tau - \bar v)^3,  
\end{equation}
and implies the mass-radius relation
\begin{equation}
\frac{ \widehat M^2}{\tau} -  \widehat R. \widehat M  + \frac{\widehat R^4}{2} = 0
\end{equation}
which can be solved for the mass 
\begin{equation}
\label{NMRE}
\widehat M = \frac{\tau \widehat R}{2} \left[1  \pm \sqrt{ 1 - \frac{2 \widehat R^2}{\tau}} \right].
\end{equation}

We remark that in (\ref{NMRE}) $\tau$ could be removed completely by a further rescaling of
$\widehat M$ and $\widehat R$. We avoid this, however, to keep the close
anlogy to the relativistic case where this is not possible. 
The behaviour of the parameters introduced above is illustrated in Table (\ref{Npar}) and 
Fig. (\ref{NMR}).

\begin{table}[h!]
\caption{The parameters of the Newtonian solutions}
\label{Npar}
\vspace*{0.5cm}

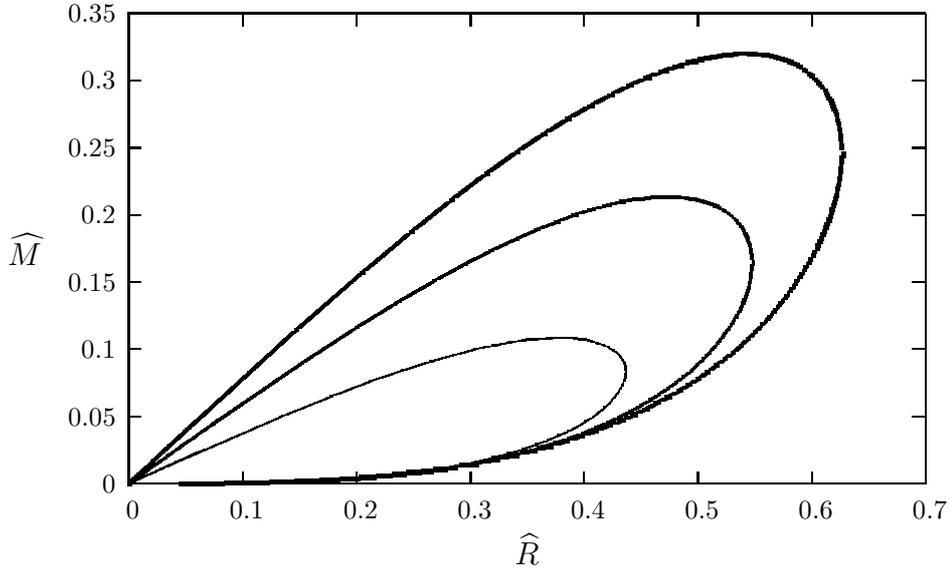
\caption{The mass-radius diagram for the Newtonian model with equation of
state (\ref{Neos}), for the values $\tau = 0.382$ (thin line), 
$\tau = 0.6$ (medium) and $\tau = 0.786$ (thick line).}
\label{NMR}     
\end{figure}

For the NEOS (\ref{Neos}) with $\rho_- \in [0,\infty)$, $\rho_+ \in (0,\infty)$
the quantities $\bar v$, $v_s$, $v_c$, $\rho_c$ and $p_c$ can take all values 
in the open interval bounded by the respective values of the "dust particle" 
and the point singularity, which are clearly unphysical themselves. These parameters are 
monotonic functions of one another, and any of them can be used to characterize the solutions.

On the other hand, the mass and the radius have extrema which follow easily from 
(\ref{NMRE}) and are also given in Table (\ref{Npar}). Fig. (\ref{NMR}) where we have
chosen the same values of $\tau$ as in Fig. (\ref{NEOS}) shows the following.
Starting with the dust particle at $\widehat R = \widehat M = v_s = 0$ 
and increasing $p_c$ and $v_s$, we follow the lower branch of the
mass-radius curve which corresponds to the minus sign in
(\ref{NMRE}). After passing the maximum radius $\widehat R = \sqrt{\tau/2}$,
the mass which is now given by the  plus sign in (\ref{NMRE}), continues to
increase till the "heaviest star" of the table is reached. Then mass and
radius drop towards the point singularity $\widehat R = \widehat M = 0$ and
$v_s =   - \tau$. The surface potential $v_s = - \widehat M/\widehat R$ 
(minus the slope of the line joining points of the curve with the origin) 
decreases monotonically along the curve, whence the latter forms precisely one "loop".

\subsection{The Relativistic Solutions}

\subsubsection{The matching}

Using Lemma A.3 of the Appendix with $\wp_{ij} = g_{ij}$, $\Phi = (1 - fV)/(1 + fV)$ and
${\cal R}_{\pm} = 512 \pi \rho_{\pm}$ we write the spherically symmetric solutions of (\ref{Rpm}) as 
\begin{eqnarray}
\label{PSV}
\frac{1 - fV}{1 + fV}  & = & \mu  \sqrt{\frac{1 + \frac{64\pi}{3}\rho_+ r^2}{1 + \frac{64\pi}{3}\mu^4 \rho_- r^2}}, \\
\label{PSg}
g_{ij} dx^i dx^j & = 
& \frac{16 \left(dr^2 + r^2 d\omega^2 \right)}{\left(1 + fV \right)^4 \left(1 + \frac{64\pi}{3}\rho_+ r^2
\right)^2}.
\end{eqnarray}

Again we have to eliminate one of the constants $f$ and $\mu$ by global conditions. 
Recall that the parameters in the EOS are now restricted by  $0 \le \rho_+ < \rho_- < \infty$ 
and so $0 \le \tau < 1$.
In the case $\rho_+ = 0$ the solutions extend to infinity 
(which can be shown independently of spherical symmetry, c.f.
\cite{WS1,WS3,MH}) 
and we set $f=1$. For $\rho_+ > 0$ the solutions are finite since $\rho_s > 0$. 
We claim that the Buchdahl solutions and the Pant-Sah solutions are given by (\ref{PSV}) and (\ref{PSg}) with
\begin{equation}
\label{Emu}
 \mu = \left\{ \begin{array}{r@{\quad \mbox{for} \quad}l} 
 {\displaystyle \frac{1}{4M} \sqrt{ \frac{3}{\pi \rho_-} } } & \rho_+ = 0,
 \\
 {\displaystyle \frac{\Sigma_+ + \Sigma_-}{2} } & \rho_+ > 0,
  \end{array} \right.
\end{equation}
and 
\begin{equation}
\label{Spm}
\Sigma_{\pm} = \tau \sqrt{(1 \pm \tau)^2 + \frac{(1 + \tau)^2 f^2 - (1 - \tau)^2}{1 - f^2}}
\end{equation}
which requires $f < 1$ to make sense. 
For $\rho_+ = 0$ equ. (\ref{Emu})  follows easily from the asymptotic condition (\ref{EM}).
 On the other hand, for $\rho_+ > 0$ the matching to Schwarzschild 
 is quite involved as the isotropic coordinates of (\ref{PSV}),(\ref{PSg}) which simplify the interior 
solutions are unsuited for the matching. We will verify (\ref{Emu}) below by matching "covariantly"
(c.f. (\ref{Esm})).

We also write the Pant-Sah solutions in the alternative form 
\begin{equation}
\label{Econf}
ds^2 = - V^2 dt^2 + \Omega^2 (dr^2 + r^2 d\omega^2)
\end{equation}
where
\begin{equation}
\label{Omega}
\Omega = \frac{4}{2\mu^2  - \Sigma^2} 
\left[ \frac{\mu^2}{(1 + fV)^2} - \frac{\tau^6}{(1 - fV)^2}\right],
\end{equation}
and the constant $\Sigma$ is given by
\begin{equation}
\label{Sigma}
 \Sigma^2 = \left\{ \begin{array}{r@{\quad \mbox{for} \quad}l} 
 {\displaystyle \mu^2} & \rho_+ = 0, \\
 {\displaystyle \frac{\Sigma_+^2 + \Sigma_-^2}{2}  } & \rho_+ > 0
  \end{array} \right.
\end{equation}
in terms of the quantities defined in (\ref{Emu}) and (\ref{Spm}).
The form (\ref{Econf}) will be useful in Section 5 for the proof of spherical symmetry of solutions with the PSEOS.
Equ. (\ref{Omega}) makes sense for $\rho_+ = 0$ as well (i.e. for the BEOS with $\rho_- \ne 0$ 
as well as for vacuum $\rho_- = 0$) and reduces to $\Omega = 4/(1 + V)^2$ in either case.

In analogy with the quantity $w$ in the Newtonian case, we now determine $W = \nabla^i V \nabla_i V$  
which is a function of $V$ in spherical symmetry. In particular, for Schwarzschild we have 
$W = (16 M)^{-2} \left(1 - V^2 \right)^4$.
For our model characterized by ${\cal R}_{\pm} = const.$,  
Lemma A.2 shows that the spherically symmetric solutions are spaces of constant curvature. 
Using the field equations (\ref{Ein}) and the general formula (\ref{cri}) with 
$\wp_{ij} = g_{ij}$, $\Phi^{\pm} = (1 \pm fV)/2$, $\widetilde \wp_{ij} = g_{ij}^{\pm}$ we have
\begin{equation}
0 = V (1 \pm V)^2  {\cal B}^{\pm}_{ij} = 
{\cal C}[(1 - f^2 V^2) \nabla_i \nabla_j V + 6 f^2 V \nabla_i V \nabla_j V].
\end{equation}

Contracting this equation with $\nabla^i V \nabla^j V$ and using the field equation 
(\ref{Alb}) and (\ref{rV}) and (\ref{pV}) gives
\begin{equation}
\frac{d}{dV} \left[ \frac{W}{(1 - f^2 V^2)^4} \right] = 
\frac{4\pi (1 - fV)}{3 f (1 + fV)}\left [ \frac{\rho_-}{(1 + fV)^2} - \frac{\rho_+}{(1 - fV)^2} \right]  
\end{equation}
which has the solution 
\begin{eqnarray} 
\label{PSW}
W & = & \frac{\pi \rho_-}{3 f^2} (1 - f^{2}V^{2})^{4}
\left[ \Sigma^2 - \frac{(1 - fV)^2}{(1 + fV)^2} - \tau^6 \frac{(1 + fV)^2}{(1 - fV)^2}\right]
=  \\
\label{Wprod}
& = & \frac{\pi \rho_- (2\mu^2 - \Sigma^2)}{12 f^2} (1 - f^2V^2)^4 \Omega \left[(1 + fV)^2 - \frac{(1 - fV)^2}{\mu^2}
\right].
\end{eqnarray}

In equ. (\ref{PSW}) $\Sigma^2$ arises as a constant of integration,
and we first verify that it is consistent with the earlier definitions 
(\ref{Sigma}), (\ref{Emu}) and (\ref{Spm}).

For $\rho_+ = 0$ this follows once again from the asymptotics, equ.(\ref{EM}).
For $\rho_+ > 0$ this is done with the "covariant" matching  condition (\ref{Egm}) 
which becomes 
\begin{equation}
\label{Esm}
 \left[ \frac{d W}{d V} - 8 \pi  \rho V \right]_{\Rightarrow \partial {\cal F}} =
\left[ \frac{d W}{d V} \right]_{ \partial {\cal V} \Leftarrow} = \frac{8 V_s W_s }{1 - V_s^2}.
\end{equation}  
Next, with a little algebra one can check (\ref{Wprod}) which contains 
 $\mu$ and $\Omega$ defined in  (\ref{Emu}) and (\ref{Omega}).
To verify that $\mu$ as defined in (\ref{Emu}) in fact agrees with
the constant appearing in (\ref{PSV}) it is simplest to use (\ref{PSW}) and the
general definition $W=\nabla_i V \nabla^i V$.

Finally we note that one can alternatively write the Pant-Sah solutions by using $V$ as a coordinate {\it everywhere}. 
(Equ. (\ref{Econf}) still contains $"r"$).
From eqs. (3.1) and (3.17) of \cite{BS2} one finds  \cite{WS1} 
\begin{equation}
ds^2 = - V^{2}dt^{2} + \frac{1}{W} dV^2 + 
\frac{9 f^2 W }{4\pi^2  \rho_-^2 (2\mu^2 -\Sigma^2)^2 (1 - f^2 V^2)^6} d\omega^2. 
\end{equation}

\subsubsection{The centre}

We now turn to the important issue of regularity at the centre. 
The latter is characterized either by $r=0$ or by the  minimum of $V$, i.e. $W =
0$. From (\ref{PSV}) and (\ref{PSg}) it is easy to see that the centre
is regular if $V > 0$; it can be made manifestly regular, i.e.
$g_{ij}dx^idx^j = (1 + O(r^2))(dr^2 + r^2 d\omega^2) $ by a suitable rescaling of $r$.

We also note that either from (\ref{PSV}) or (\ref{Wprod}) it follows that regularity is equivalent to $\mu < 1$.
For $\rho_+ = 0$ this entails the {\it lower bound} $M^2  > 3/16\pi\rho_-$ {\it for the
mass}, while there is no upper bound.
For $\rho_+ > 0$ we have collected  in Table (\ref{Epar1}) the most
important parameters which are monotonic functions of one another and 
provide unique characterizations of the model.
We use the shorthand ${\cal T}_{\pm} = \sqrt{1 \pm \tau + \tau^2}$.

\begin{table}[h]
\caption{Some parameters for the Pant-Sah solutions}
\label{Epar1}
\vspace*{0.5cm}
\begin{tabular}{|r||c|c|c|c|c|}
{} & $f$ & $V_s$ & $V_c$ & $\rho_c$ & $p_c$ \\ [1.0ex] \hline  \hline
{} & {} & {} & {} & {} & {} \\ 
dust particle & {\large $\frac{1 - \tau}{1 + \tau}$ }  & $1$ & $1$ & 
{\large $\frac{32 \rho_- \tau^5}{1 + \tau^4}$ } & $0$   \\ [1.5ex] 
singular centre &  {\large $\frac{(1 - \tau){\cal T}_+^2}{(1 + \tau){\cal T}_-^2}$ }  &  
{\large $\frac{{\cal T}_-^2}{{\cal T}_+^2}$} & $0$ & $ \rho_- + \rho_+$ & $\infty$  
\end{tabular}   

\end{table}

The allowed parameter values are bounded by the respective ones of the "dust particle" 
with $p_c = 0$ and $V_c = V_s = 1$, and the model with singular centre for which
$p_c = \infty$ and  $V_c = 0$.
 Like their Newtonian counterparts these limits are unphysical, but unlike
the Newtonian ones the singular solution has now finite extent.
The dust particle has $\rho_c = \rho_s > 0$, while as the singular centre is approached
 $\rho_c$ always stays finite (in contrast to the Newtonian case) due to the 
"stiffness" of the PSEOS. This singular model also has largest redshift,
 which can be tested against the Buchdahl limit \cite{HB2} 
$V_s \ge 1/3$. The latter is saturated for fluids of constant density
only. Such fluids are approached by the present models for $\tau \rightarrow 1$ (c.f.
Equ.(\ref{rhoc})), and in fact we find that $V_s \rightarrow 1/3$ in this limit.

\subsubsection{The mass-radius relation}

To obtain mass and radius we use  $1 - V_s^2 = 2M/R$, (\ref{Esm}) and (\ref{PSW}) at the surface. 
In terms of the rescaled variables (\ref{resc}) we obtain

\begin{eqnarray} 
\label{mass}
\widehat R^2 & = &  
\frac{1 - f^{2}}{4 \tau} \left[(1 + \tau)^2 - \frac{(1 - \tau)^{2}}{f^2} \right], \\
 \widehat M^2  & = & 
\frac{1 - f^{2}}{16 \tau (1 + \tau)^4} \left[ (1 + \tau)^2 - \frac{(1 - \tau)^{2}}{f^2} \right]^{3}. 
\end{eqnarray}

Eliminating $f$ gives the mass-radius relation
\begin{equation}
\frac{(1 + \tau)^2}{2\tau} \widehat M^2 - (1 + \widehat R^2)\widehat R \widehat M + \frac{\widehat R^4}{2}
= 0
\end{equation}

which can be solved for the mass

\begin{equation}
\label{EMR}
\widehat M = \frac{\tau \widehat R}{(1 + \tau)^2}  
\left[1 + \widehat R^2 \pm  \sqrt{\left(\tau - \widehat R^2 \right )\left(\frac{1}{\tau} - \widehat R^2 \right)}\right].
\end{equation}

The extrema of mass and radius are listed in Table (\ref{Epar2}).
(Recall that ${\cal T}_{\pm} = \sqrt{1 \pm \tau + \tau^2}$).

\begin{table}[h!]
\caption{Surface potential, radius and mass of the Pant-Sah solution}
\label{Epar2}
\vspace*{0.5cm}


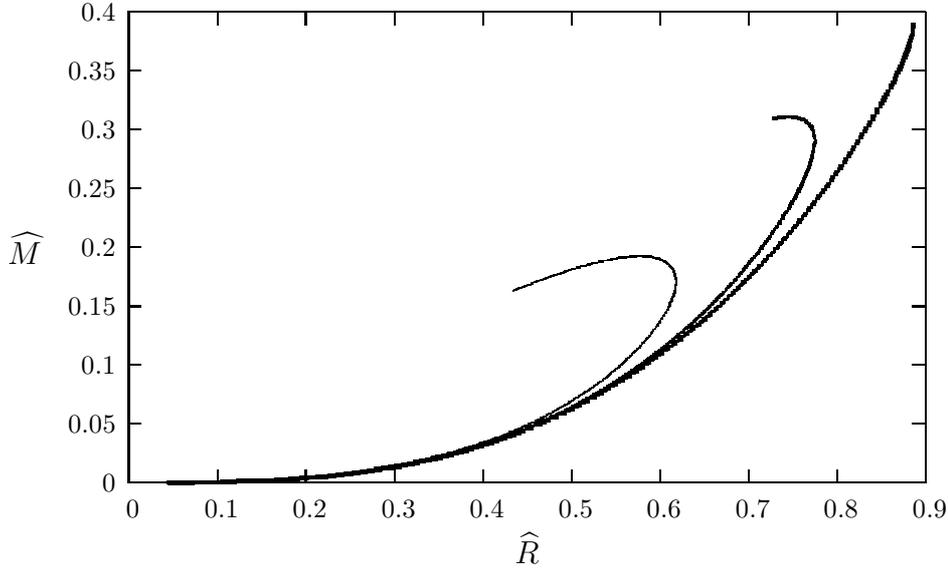
\caption{The mass-radius diagram for the Pant-Sah solution for the values $\tau=0.382$
(thin line), $\tau = 0.6$ (medium) and $\tau=0.786$ (thick line).}
\label{PSMR}      
\end{figure}

As in the Newtonian case the surface potential characterizes the solution uniquely,
which implies the loop-like structure of the mass-radius curves, Fig.
(\ref{PSMR}).
We first describe the diagrams for sufficiently small values of $\tau$
such as $\tau = (3 -\sqrt{5})/2 \approx 0.382$ and $\tau = 0.6$.
(These particular values correspond to $V_s^2 = 1/3$ and $V_s^2 = 1/6$ at
the respective mass maxima).  
Starting with the dust particle and increasing $p_c$, $V_s$ decreases and  
the mass-radius curve corresponds to the minus sign in front of the root in (\ref{EMR}). 
At the  maximum radius which is now at $\widehat R = \sqrt{\tau}$,
we pass to the plus sign. From then onwards the star shrinks,
reaching its maxium mass at some lower value of $V_s$, and subsequently
losing weight as well. 
In contrast to the Newtonian case, the singular model now prevents the
"mass-radius loop" from closing. As already mentioned in the discussion of
Table (\ref{Epar1}), at some finite size of the star the central pressure
diverges, and this is where the curves in Fig. (\ref{PSMR}) terminate.

For $\tau=0.6$, the star with maximal mass still has a regular centre. 
However, for larger values of $\tau$, the central pressure diverges 
before the mass maximum or even the maximal radius are reached. 
This means that the  "biggest star" and the "heaviest star" in Table (3) only make sense 
if the respective values of $V_s$ are larger than the ones given for the
"singular centre". For $\tau = [(\sqrt{5} - 1)/2]^{\frac{1}{2}} \approx 0.786$ 
the star with maximal radius is precisely the first one with singular centre,
and the meaningful part of the mass-radius curve is monotonic (c.f. Fig. (\ref{PSMR})).

We finally note that the mass radius curves for the "softer" PSEOS (such as $\tau = 0.382$
in Fig (\ref{PSMR})) resemble strikingly the mass-radius curves for 
quark stars \cite{WNR}-\cite{DBDRS} 
(in particular those with "harder" equations of state such as \cite{RBDD}, \cite{DBDRS}).
Putting  $\rho_- = 3~\mbox{GeV/fm}^3$ in the PSEOS with  $\tau = 0.382$ 
one obtains typical values of about 1.5 solar masses for the maximal mass 
and about 7 km for the maximal radius.    
However, this coincidence should not be overestimated.
As mass and radius are obtained by integration, they "smooth out" differences in the
EOS, which seem quite substantial even at moderate densities.
Moreover, we recall that for the PSEOS the pressure always diverges at finite density, 
whereas the EOS of "ultrarelativistic" quarks is nowhere too far from $p = \rho/3$. 
This discrepancy prevents us from modelling extreme quark conditions
and has rather drastic consequences for the mass-radius relation.
As follows from Harrison et. al. \cite{HTWW} and has been shown rigorously 
by Makino \cite{TM}, if the quotient $p(\rho)/\rho$ for some given EOS tends to a constant 
sufficiently fast for $\rho \rightarrow \infty$ or $p \rightarrow \infty$, 
the mass-radius curve develops the form of a "spiral", with an infinite number
of twists, for high central pressure.
While the EOS for quark stars given in the literature seem safely 
within the range of the Makino theorem, the mass-radius diagrams are normally  
not drawn till the spiral sets on. On the other hand, the mass-radius diagrams for the
Pant-Sah solutions are single open loops, 
which we have drawn to the end (infinite central pressure) in Fig. (\ref{PSMR}).

\section{Proofs of Spherical Symmetry}

In Newtonian theory Lichtenstein \cite{Lich} has given a proof  
of spherical symmetry of static perfect fluids which satisfy 
$\rho \ge 0$ and $p \ge 0$. 
Under the same condition on the equation of state, Masood-ul-Alam
has recently proven spherical symmetry in the relativistic case 
by using a substantial extension of the positive mass theorem \cite{MA1}.
For the relativistic model considered in this paper, spherical symmetry
 is a consequence of the uniqueness theorem of Beig and Simon \cite{BS2}. 
In Sect. 4.2  we reproduce the core of this proof for the present model,
for which it simplifies.

In Sect. 4.1 we give a version of the Newtonian proof which resembles as closely as
possible the relativistic proof, substituting the positive mass theorem
by the virial theorem. A proof along the same lines has been sketched in
\cite{BS2} for fluids of constant density.

\subsection{The Newtonian Case}

We use the notation of sections 2.1.1 and 2.1.2 with the following
additions and modification.
We define $w = g^{ij}\nabla_i v\nabla_j v$ as in Sect. 2.1.1.
However, for a given model described  by $v_s$, 
we now denote by  $w_0(v)$ the function of $v$ and $v_s$ 
defined by the r.h. side equ. (\ref{wv}). 
Note that this function may become negative, which happens if the central potential 
$v_c$ of the given model is less than the central potential of the spherical symmetric 
model with the same $v_s$. The proof of spherical symmetry proceeds by
showing that  $w$ and $w_0$ coincide. We split this demonstation into a series
of Lemmas. \\ \\
{\bf Lemma 4.1.1. (The virial theorem)}\\ 
 For all static asymptotically flat fluids as described in Sect. 2.1.1., we
have, denoting the volume element by $d\eta$,
\begin{equation}
\label{vir}
\int_{\cal F} (6p + \rho v) d\eta = 0.
\end{equation} \\ 
{\bf Remark.} In kinetic gas theory, the two terms in the integral (\ref{vir})
are four times the kinetic energy and twice the potential energy of the
particles, respectively.\\ \\
{\bf Proof.}  Let $\xi_i$ be a dilation in flat space, i.e. $\nabla_{(i}\xi_{j)} =
g_{ij}$. (In cartesian coordinates, $g_{ij} = \delta_{ij}$ and $\xi_i =
x^i$). Let $v$, $\rho$ and $p$ define an asymptotically flat  
Newtonian model as in Sect. 2.1.1.. 
Then there holds the Pohozaev identity \cite{SP}
\begin{equation}
\label{Poho}
\nabla_i \left [\left(\xi^j \nabla_j v + \frac{1}{2} v \right) \nabla^i v - 
 \frac{1}{2} w \xi^i + 4\pi p \xi^i \right] = 2 \pi (6p + \rho v)
\end{equation}
which is verified easily. We integrate this equation over ${\cal M}$ and apply the divergence
theorem, using that the integrand in brackets on the left is continuous at the
surface. Due to the asymptotic conditions (\ref{NM}), (\ref{Nas}) the boundary integral at
infinity vanishes which gives the required result (\ref{vir}). 
\hfill $\Box$ \\ \\
{\bf Lemma 4.1.2.} 
For solutions with the NEOS (\ref{Neos}), we have
\begin{eqnarray}
\label{PSvir}
0 & = & \int_{\cal F} (6p + \rho v) d\eta = - \rho_+ Y + \bar v M \\
\label{intw}
0 & = & \int_{\cal F} \frac{w - w_0}{(\bar v - v)^4} d\eta 
\end{eqnarray}
where $Y = \int_{\cal F} d\eta$ is the volume of the fluid.
In particular,  $\rho_+ = 0$ iff the solutions extend to infinity ($0 = v_s = \bar v - \tau$) \\ \\
{\bf Proof.} For the Newtonian model the virial theorem (\ref{vir}) and (\ref{rhop}) imply
\begin{equation}
\label{PSvirp}
0 = \int_{\cal F} (6p + \rho v) d\eta = 
\int_{\cal F} [6p - \rho (\bar v - v)] d\eta + \bar v \int_{\cal F}  \rho d\eta = - \rho_+ Y + \bar v M
\end{equation} 
which proves (\ref{PSvir}). To show (\ref{intw})
we use the divergence theorem, (\ref{Poi}), (\ref{rhop}),
(\ref{Nsurf}) and (\ref{wv}). We obtain
\begin{eqnarray}
& & 3 \int_{\cal F} \frac{w - w_0}{(\bar v - v)^4} d\eta  = 
\int_{\cal F} \left [ \nabla_i \frac{\nabla^i v}{(\bar v - v)^3}  
- \frac{\Delta v}{(\bar v - v)^3}  - \frac{3 w_0}{(\bar v - v)^4} \right] d\eta =
\nonumber \\
\label{ww0}
{} & = &  \frac{1}{\tau^3} \int_{\partial {\cal F}} \nabla_i v~ d{\cal S}^i
-  \frac{4\pi \rho_- \tau^3}{\bar v} \int_{\cal F} d\eta =
\frac{4\pi}{\tau^3} \left( M - \frac{\rho_+ Y}{\bar v} \right)  
\end{eqnarray}
Equ. (\ref{PSvir}) and (\ref{ww0}) now give the required result. 
\hfill $\Box$ \\ \\
{\bf Lemma 4.1.3.} For the NEOS there holds, inside ${\cal F}$
\begin{equation}
\label{NLapf}
 2 \Delta^- ~\frac{ w -  w_0}{(\bar v - v)^4} = (\bar v - v)^2  {\cal B}_{ij}^- {\cal B}^{ij}_- \ge 0.
\end{equation}
\\ {\bf Proof.}
We first note that formula (\ref{cri}) with $\wp_{ij} = g_{ij}^-$, $\Phi = (\bar v - v)^{-1}$,
 and  $\widetilde \wp_{ij} = \Phi^{4} \wp_{ij} = g_{ij}$ implies  
$ {\cal B}_{ij}^- = - (\bar v - v)^{-2}  {\cal C}_- [ \nabla_i^- \nabla_j^- (\bar v - v)^2]$.
If follows that
\begin{eqnarray}
 2 \Delta^- ~ \frac{w -  w_0}{(\bar v - v)^4} & = & 
- \nabla^i_- [ {\cal B}_{ij}^- \nabla^j_- (\bar v - v)^2] =
 (\bar v - v)^2  {\cal B}_{ij}^- {\cal B}^{ij}_- - \nonumber \\
 & - & \frac{1}{6} [\nabla^i_- (\bar v - v)^2 ] \nabla_i^- {\cal R}^- = 
(\bar v - v)^2  {\cal B}_{ij}^- {\cal B}^{ij}_- 
\end{eqnarray}
where we have used the Bianchi identity $\nabla^i_- {\cal B}_{ij}^- = \nabla_j^- {\cal R}^-/6$ 
and the fact that ${\cal R}^- = const. $ for the NEOS.
\hfill $\Box$ \\ \\
{\bf Lemma 4.1.4.} In ${\cal V}$ we have 
\begin{equation}
\label{NLapv}
\Delta^- ~ \frac { w -  w_0}{|v|^3 (\bar v - v)} =  \frac{|v|^7}{(\bar v - v)^5} \widehat{\cal B}_{ij} 
\widehat {\cal B}^{ij} \ge 0
\end{equation}
where $\widehat{\cal B}_{ij}$ is the trace-free part of the Ricci tensor
w.r. to the metric $\widehat g_{ij} = v^4 g_{ij}$.  \\ \\
{\bf Proof.} 
In the vacuum region (\ref{NLapf}) still holds (we set $\bar v = 0$), since the metric
$\widehat g_{ij}$ has curvature $\widehat {\cal R} = 0$.
It follows that
\begin{eqnarray}
 \Delta^- ~ \frac { w -  w_0}{|v|^3 (\bar v - v)} & = &
\frac{v^6}{(\bar v - v)^6}  
\widehat \nabla_i \left [ \frac{(\bar v - v)^2}{v^2} 
\widehat \nabla^i ~ \left( \frac{|v|}{\bar v - v}~  \frac{w - w_0}{v^4} \right) \right]
= \nonumber \\
& = & \frac{|v|^5}{(\bar v - v)^5}  \widehat \Delta~ \frac{w - w_0}{v^4} =  
 \frac{|v|^7}{(\bar v - v)^5} \widehat {\cal B}_{ij} \widehat {\cal B}^{ij}
\end{eqnarray}
where $\widehat \nabla^i$ and $\widehat \Delta$ refer to $\widehat g_{ij}$.  
\hfill $\Box$ \\ \\
{\bf Lemma 4.1.5.} On ${\cal M} = {\cal V} \cup {\cal F}$, we obtain $w \le w_0$.
\\ \\
{\bf Proof.}
The weak maximum principle applied to (\ref{NLapf}) on ${\cal F}$
implies that $(w - w_0)/(\bar v - v)^4$ takes on its maximum at some point
$q \in \partial {\cal F}$, i.e.
\begin{equation}
\label{Nin2}
\sup_{\cal F} \frac{w - w_0}{(\bar v - v)^4} \le \max_{\partial {\cal F}} \frac{w - w_0}{(\bar v - v)^4}
= \left. \frac{w - w_0}{(\bar v - v)^4} \right|_q
\end{equation}
On the other hand, the weak maximum principle applied to (\ref{NLapv}) shows that either 
$(w - w_0)/|v|^3 (\bar v - v)$ takes on its (absolute) maximum at infinity (where it vanishes)
or that it has a positive maximum on $\partial {\cal F}$.
In the latter case  this maximum is located at $q$
since $v$ is constant on $\partial {\cal F}$. This leads to a contradiction
as follows. Taking $n^i$ to be the normal to $\partial {\cal F}$ directed towards
infinity, we have 
\begin{equation}
\label{Nbd}
\left. n^{i} \nabla_i~ \frac{w - w_0}{(\bar v - v)^4} \right|_q = 
\left. \frac{|v_s|^3}{\tau^3}  n^{i} \nabla_i ~ \frac{w - w_0}{(\bar v - v) |v|^3} \right|_q
+ 3 \frac{\bar v (w - w_0)}{\tau^3 |v_s|^3} \left. n^i \nabla_i~ |v| \right|_q.
\end{equation} 
By the boundary point lemma \cite{GT} applied to (\ref{NLapv}), 
the first term on the right of (\ref{Nbd}) is negative,
and the same applies to the second term by virtue of $\Delta v = 0$.
It follows that
\begin{equation}
\label{Nin1}
\left. n^{i} \nabla_i ~\frac{w - w_0}{(\bar v - v)^4} \right|_q  < 0
\end{equation}
But as $\bar v -  v$ is $C^1$ on ${\cal M}$, this contradicts
(\ref{Nin2}). Hence we are left with  $(w - w_0)/|v|^3 (\bar v - v) \le 0$ in ${\cal V}$ which, 
together with (\ref{Nin2}) implies $(w - w_0)/(\bar v - v)^4 \le 0$
everywhere on ${\cal M}$.
This proves the Lemma. \hfill $\Box$ \\ \\
{\bf Lemma 4.1.6}. On ${\cal M} = {\cal V} \cup {\cal F}$, we have $w = w_0$.
Furthermore $w = w_0(v)$ is positive for $v > v_{min}$, smooth in $[v_{min}, v_s)$ 
and such that at $v_{min}$ there holds $w_0 = 0$ and $ dw/dv \ne 8\pi \rho$.
\\ \\
{\bf Proof.} The first assertion follows from Lemmas 4.1.2 and 4.1.5.,
and the rest is easily checked. \hfill $\Box$ \\ \\
{\bf Proposition 4.1.7.}  Asymptotically flat solutions with the NEOS are spherically symmetric
and uniquely defined by $w_0$. \\ \\
{\bf Proof.} A trivial modification of a relativistic result, Lemma 4 of \cite{BS2}, 
has the conclusion of Lemma 4.1.6. as hypothesis. The conclusion of this
modified Lemma is the proposition.

\subsection{The Relativistic Case} 

We use the notation of the previous sections with modifications analogous to the Newtonian case. 
We recall that $W = g^{ij}\nabla_i V \nabla_j V$, and for a given model described by $V_s$ 
we now denote by $W_0(V)$ the function of $V$ and $V_s$ defined by the r.h. side equ. (\ref{PSW}). 
Again this function becomes negative if the central potential $V_c$ of the given model is less than the 
central potential of the spherical symmetric model with the same $V_s$.
We first prove that $W$ and $W_0$ coincide \cite{Lind},
which is done in a series of Lemmas. From Lemma 4.2.3. onwards, they are direct counterparts 
of the Newtonian ones in the previous section. \\ \\
{\bf Lemma 4.2.1. (The vanishing mass theorem)}.
We recall that an asymptotically flat Riemannian manifold with non-negative scalar curvature 
and vanishing mass is flat \cite{SY}. \\ \\
{\bf Lemma 4.2.2.} For fluids with PSEOS, we have 
\begin{equation}
\label{Rstar}
{\cal R}_{\star} = \frac{1536 \tau^6 \mu^2 f^2 (W_0 - W)}{(2\mu^2 - \Sigma^2)(1 - f^2V^2)^4}
\end{equation}
where ${\cal R}_{\star}$ is the scalar curvature w.r.to the metric
 $g_{ij}^{\star}  = \Omega^{-2} g_{ij}$, with $\Omega(V)$ defined in (\ref{Omega}).
For $\rho_+ = 0$, i.e. for the BEOS and for vacuum, we obtain ${\cal R}_{\star} \equiv 0$.
\\ \\
{\bf Proof.}
For the curvature w.r.t. to $g_{ij}^{\star}$ we obtain  
\begin{eqnarray}
{\cal R}_{\star} & = & 2 \left[3 \left(\frac{d\Omega}{dV} \right)^2 - 2 \Omega \frac{d^2
\Omega}{dV^2} \right] 
(W_0 - W) =  \nonumber\\
& = & 16\pi\Omega^2 \left[ \rho + (\rho + 3p) \frac{V}{\Omega} \frac{d\Omega}{dV}\right]
\left( 1 - \frac{W}{W_0} \right).
\end{eqnarray}
Here the first equation holds for  conformal rescalings 
of the form $g_{ij}^{\star}  = \Omega^{-2} g_{ij}$ (for any $g_{ij}$ and
$\Omega(V)$ if  ${\cal R}$ is a function of $V$ only)  
while the second one uses the general formula (\ref{csc}), property (\ref{Econf}), 
and the field equations (\ref{Alb}) and (\ref{Ein}). 
Now (\ref{Rstar}) follows by using the explicit forms (\ref{rV}), (\ref{pV}) and (\ref{Omega}) 
of $\rho$, $p$ and $\Omega$. \\ \\
{\bf Lemma 4.2.3.} 
For solutions with PSEOS, we have
\begin{equation}
\label{ELapf}
 \Delta^{\prime} ~ \frac{W - W_0}{(1 - f^2V^2)^4}  = 
 \frac{V^4 (1 \pm fV)^2}{(1 \mp fV)^{10}}  {\cal B}_{ij}^{\pm} {\cal B}^{ij}_{\pm} \ge 0
\end{equation}
where $\Delta^{\prime}$ refers to the metric $g_{ij}^{\prime} = (1 - f^2V^2)^4 g_{ij}/16 V^2$.
\\ \\
{\bf Proof.}
We first note that formula (\ref{cri}) with $\wp_{ij} = g_{ij}^{\pm}$, $\Phi^{\pm} = 2/(1 \pm fV)$,
and  $\widetilde \wp_{ij} = \Phi^{4}_{\pm} \wp_{ij} = g_{ij}$ implies, 
together with the field equation (\ref{Ein}),
\begin{equation}
\label{EB}
{\cal C}^{\pm}\left[ \nabla_i^{\pm} X_j^{\pm} \right ] =  \alpha^{\pm} {\cal B}_{ij}^{\pm}
\end{equation}
where we have defined
\begin{equation}
X_i^{\pm} = \frac{1 \pm fV}{(1 \mp fV)^3}  \nabla_{j}V  
\qquad \alpha^{\pm} = \frac{V (1 \pm fV)^{2}}{(1 \mp fV)^4}.
\end{equation}
Then we find from (\ref{EB}) that
\begin{eqnarray}
 \frac{(1 \mp fV)^6}{V^3} \Delta^{\prime}~ \frac{W - W_0}{(1 - f^2V^2)^4} 
& = &
 \nabla^i_{\pm} \left[{\cal B}_{ij}^{\pm} X^j_{\pm} \right] =
 \alpha^{\pm}  {\cal B}_{ij}^{\pm} {\cal B}_{ij}^{\pm}  + \nonumber \\
& + & \frac{1}{6} X^i_{\pm} \nabla_i^{\pm} {\cal R}^{\pm} =  
\alpha^{\pm} {\cal B}_{ij}^{\pm} {\cal B}^{ij}_{\pm} 
\end{eqnarray}
where we have used the Bianchi identity $\nabla^i_{\pm} {\cal B}_{ij}^{\pm} = \nabla_j^{\pm}
{\cal R}^{\pm}/6$  and the fact that ${\cal R}^{\pm} = const. $ for our model.
\hfill $\Box$ \\ 

Note that the argument of the Laplacian on the l.h. side of (\ref{ELapf}) agrees with ${\cal R}_{\star}$ 
as given in (\ref{Rstar}) modulo a constant factor. In other words, (\ref{Rstar}) and (\ref{ELapf}) 
show that $\Delta^{\prime} {\cal R}_{\star} \ge 0$. \\ \\
{\bf Lemma 4.2.4.} In ${\cal V}$, we have
\begin{equation}
\label{ELapv}
\Delta^{\prime}~\frac{W - W_0}{(1 - V^2)^3 (1 - f^2 V^2)} = 
\frac{V^3 (1 \pm V)^7}{(1- f^2 V^2)^5 (1 \mp V)^5} 
\widehat {\cal B}_{ij}^{\pm} \widehat {\cal B}^{ij}_{\pm} \ge 0
\end{equation} 
where $\widehat {\cal B}_{ij}^{\pm}$ are the trace free parts of the Ricci
tensors w.r. to the metrics $\widehat g_{ij}^{\pm} = (1 \pm V)^4 g_{ij}/2$.\\ \\
{\bf Proof.} In vacuum (\ref{ELapf}) still holds (we set $f=1$), since the
metrics $\widehat g_{ij}^{\pm}$ have vanishing curvatures $\widehat {\cal R}^{\pm}$.  
It follows that
\begin{eqnarray}
 & & \Delta^{\prime}~\frac{W - W_0}{(1 - V^2)^3 (1 - f^2 V^2)} = \nonumber \\
& = & \frac{(1 - V^2)^6}{(1 - f^2V^2)^6} \widehat \nabla_i^{\pm} \left[\frac{(1 - f^2 V^2)^2}{(1 - V^2)^2}
\widehat \nabla^i~ \left( \frac{1 - V^2}{1 - f^2 V^2} ~ \frac{W - W_0}{(1 - V^2)^4} \right) \right] = \nonumber \\ 
& = & \frac{(1 - V^2)^5}{(1 - f^2 V^2)^5} \widehat \Delta^{\pm}~ \frac{W - W_0}{(1 - V^2)^4} =
\frac{V^3 (1 \pm V)^7}{(1- f^2 V^2)^5 (1 \mp V)^5} \widehat {\cal B}_{ij}^{\pm} \widehat {\cal B}^{ij}_{\pm}.
\end{eqnarray} 
where $\widehat \nabla^{\pm}$ and $\widehat \Delta^{\pm}$ refer to $\widehat g_{ij}^{\pm}$, respectively.
\hfill $\Box$ \\ \\
{\bf Lemma 4.2.5.} On ${\cal M} = {\cal V} \cup \partial {\cal F} \cup {\cal F}$, we have $W \le W_0$. \\ \\ 
{\bf Proof.} 
The weak maximum principle applied to (\ref{ELapf}) on ${\cal F}$ implies that $(W - W_0)/(1 - f^2 V^2)^4$ 
takes on its maximum on some point $q \in \partial {\cal F}$, i.e.
\begin{equation}
\label{Ein2}
\sup_{\cal F} \frac{W - W_0}{(1 - f^2 V^2)^4} \le \max_{\partial {\cal F}} \frac{W - W_0}{(1 - f^2 V^2)^4}
= \left. \frac{W - W_0}{(1 - f^2 V^2)^4} \right|_q
\end{equation}
On the other hand, the weak maximum principle applied to (\ref{ELapv}) shows that either
$(W - W_0)/(1 - V^2)^3 (1 - f^2 V^2)$ takes on its (absolute) maximum at infinity (where it vanishes)
or that it has a positive maximum at some point $q \in \partial {\cal F}$.
In the latter case this maximum is located at $q$,
as $V$ is constant on $\partial {\cal F}$. This leads to a contradicition as
follows. Taking $n^i$ to be the normal to $\partial {\cal F}$ directed
towards infinity, we have 
\begin{eqnarray}
\label{Ebd}
\left. n^{i} \nabla_i~ \frac{W - W_0}{(1 - f^2 V^2)^4} \right|_q & = &
\left. \frac{(1 + \tau)^2 (1 - V_s^2)^3}{4\tau}  n^{i} \nabla_i ~ \frac{W - W_0}{(1 - f^2 V^2)(1 - V^2)^3} \right|_q
- \nonumber \\
& -& \frac{3 (1 + \tau)^6 (1 - f^2) V_s (W - W_0)}{32 \tau^3 (1 - V_s^2)^3} \left. n^i \nabla_i~ V \right|_q
\end{eqnarray} 
By the boundary point lemma applied to (\ref{ELapv}), the first term on the right of (\ref{Ebd}) is negative,
while the second term (without the minus in front) is positive by virtue of $\Delta V = 0$.
It follows that
\begin{equation}
\label{Ein1}
\left. n^{i} \nabla_i ~\frac{W - W_0}{(1 - f^2 V^2)^4} \right|_q  < 0
\end{equation}
But as $1 - f^2 V^2$ is $C^1$ on ${\cal M}$, this contradicts
(\ref{Ein1}). Hence we are left with  $(W - W_0)/(1 - V^2)^3 (1 - f^2 V^2) \le 0$ in ${\cal V}$ which, 
together with (\ref{Ein2}) implies $(W - W_0)/(1 - f^2 V^2)^4 \le 0$
everywhere on ${\cal M}$.
This proves the Lemma. \hfill $\Box$ \\ \\
{\bf Lemma 4.2.6}. On ${\cal M} = {\cal V} \cup {\cal F}$ we have $W = W_0$.
Furthermore $W = W_0(v)$ is positive
for $V > V_{min}$, smooth in $[V_{min}, V_s)$ and such that at $V_{min}$
there holds  $W_0 = 0$ and $ dW/dV \ne 8\pi (\rho + 3p) V$.
\\ \\
{\bf Proof.} The first assertion follows from Lemmas 4.2.2 and 4.2.5.,
and the rest is easily checked. \hfill $\Box$. \\ \\
{\bf Proposition 4.2.7.}  Asymptotically flat solutions with the PSEOS are spherically symmetric
and uniquely defiend by $W_0$. \\ \\
{\bf Proof.} Lemma 4 of \cite{BS2} (which is a reformulation of 
results of \cite{AA,HK}) has the conclusion of Lemma 4.2.6 as hypothesis.
The conclusion of the former Lemma is the proposition.
 \\ \\
{\large\bf Acknowledgements.}
I am grateful to J. Mark Heinzle, Kayll Lake, Marc Mars and to the referees for useful comments
on the manuscript.

\section{Appendix}

We recall here the basic formulas for the behaviour of the curvature
under conformal rescalings of the metric and the standard form of metrics
of constant curvature. We give the proofs of the latter two Lemmas. \\ \\ 
{\bf Lemma A.1.} 
Let $\wp_{ij}$, $\widetilde \wp_{ij} = \Phi^4 \wp_{ij}$ be conformally related metrics on a
3-dimensional manifold ${\cal M}$.
Then the scalar curvatures  $\Re$, $\widetilde \Re$ and
the trace-free parts  ${\cal B}_{ij}= {\cal C}[\Re_{ij}]$ and 
$\widetilde {\cal B}_{ij} = \widetilde {\cal C}[\widetilde \Re_{ij}]$ 
of the Ricci tensors $\Re_{ij}$ and $\widetilde \Re_{ij}$ behave as
\begin{eqnarray}
\label{csc}
 - \frac{1}{8} \widetilde \Re \Phi^5 & = & \left(\Delta - \frac{1}{8} \Re \right) \Phi
\\
\label{cri}
\widetilde {\cal B}_{ij} & = & {\cal B}_{ij} - 2\Phi^{-1} {\cal C}[\nabla_i\nabla_j \Phi] +  6
\Phi^{-2} {\cal C}[\nabla_i \Phi \nabla_j \Phi]
\end{eqnarray}
where $\nabla_i$ is the gradient and $\Delta = \wp^{ij}\nabla_i \nabla_j $ 
the Laplacian of $\wp_{ij}$.\\ \\
{\bf Lemma A.2.} Any smooth, spherically symmetric metric $({\cal M}, \wp_{ij})$
with constant scalar curvature $\Re$ is a space of constant curvature
(i.e. ${\cal B}_{ij} = 0$) and can be written as
\begin{equation}
\label{cc}
ds^2 = \frac{1}{\left(1 + \frac{1}{24} \Re r^2 \right)^2}
(dr^2 + r^2 d\omega^2)
\end{equation}
{\bf Proof.} By solving ODEs we can write $\wp_{ij}$ in isotropic coordinates
as
\begin{equation}
\label{iso}
ds^2 = \Phi(r)^4 (dr^2 + r^2 d\omega^2)
\end{equation} 
To determine $\Phi$, we solve (\ref{csc}) with $\wp_{ij}$ flat and
$\widetilde \Re = const.$. Near the center, the solution determined uniquely
 by the initial values $\Phi_c = 1$ and $\partial \Phi/\partial x^i|_c = 0$ reads 
\begin{equation}
\label{Phi}
\Phi(r) =   \frac{1}{\sqrt{1 + \frac{1}{24} \Re r^2}}
\end{equation}
The solution can be extended analytically as far as required, which gives (\ref{cc}). \hfill $\Box$ \\ \\
{\bf Lemma A.3.} Let  $({\cal M}_+, \wp_{ij}^+)$ and $({\cal M}_-, \wp^-_{ij})$ be 3-dimensional,  
spherically symmetric manifolds with smooth metrics and with constant scalar curvatures $\Re_{+}, \Re_-$. 
Then the smooth, spherically symmetric solutions $\Phi_+(r)$ to the equation (\ref{csc}) 
on $({\cal M}_+, \wp^+_{ij})$ are given by
\begin{equation}
\label{Phi}
\Phi_+(r) =    \mu \sqrt{\frac{1 + \frac{1}{24}\Re_+ r^2}{1 + \frac{1}{24}\mu^4 \Re_- r^2}}
\end{equation}
where $\mu$ is a constant. \\ \\
{\bf Proof.}
Using Lemma A.2., the required solution is determined by a conformal rescaling 
$\wp_{ij}^- = \Phi_+^4 \wp_{ij}^+$ between spaces of constant curvature. 
Writing the metrics in the forms (\ref{cc}) gives 
\begin{equation}
\label{cc+-}
 \frac{\Phi_+(r_+)^4} {\left(1 + \frac{1}{24}\Re_+ r_+^2 \right)^2}(dr_+^2 + r_+^2
d\omega^2) = 
\frac{1} {\left(1 + \frac{1}{24}\Re_- r_-^2 \right)^2}(dr_-^2 + r_-^2 d\omega^2) 
\end{equation}
Hence $\Psi_+ (r_+)$ defined by
\begin{equation}
\Psi_+(r_+) = \Phi_{+}(r_+) \sqrt{\frac{1 + \frac{1}{24} \Re_- r_-^2}
{1 + \frac{1}{24} \Re_+ r_+^2}}
\end{equation}
with a yet unknown relation $r_- = r_-(r_+)$, 
determines a conformal diffeomorphism of flat space, i.e.
\begin{equation}
\label{cd}
 \Psi_+^4 (dr_+^2 + r_+^2 d\omega^2) = (dr_-^2 + r_-^2 d\omega^2) 
\end{equation}

By (\ref{csc}) all such diffeomorphisms are solutions of  $\Delta \Psi_+ = 0$ on flat space, 
and hence given by $\Psi_+ = \mu + \nu/r_+$ for some constants $\mu$ and $\nu$. 
A consistency check with (\ref{cd}) now shows that either $\nu = 0$ and $r_-
= \mu^2 r_+$ or $\mu = 0$ and $r_- = \nu^2/r_+$.  Setting $r_+ = r$, the
first case leads directly to (\ref{Phi}) while in the second case we have to
put $\mu^2 = 24/(\nu^2 \Re_-)$.  The solution again extends analytically as far
as needed. \hfill $\Box$



\end{document}